\documentclass[nofootinbib,aps,showpacs,11pt]{revtex4}
\usepackage{graphicx}
\usepackage{cancel}
\usepackage{amssymb}
\usepackage{textcomp}
\usepackage{amsmath}
\usepackage{bm}
\usepackage{times}
\usepackage{epsfig}
\usepackage{color}
\usepackage{axodraw}
\usepackage{subfig}

\def\lsim{\:\raisebox{-0.5ex}{$\stackrel{\textstyle<}{\sim}$}\:}

\def \ee{\end{equation}}
\def \be{\begin{equation}}
\def \bea{\begin{eqnarray}}
\def \eea{\end{eqnarray}}

\begin{document}

\title{\Large Gravitino Dark Matter and Neutrino Masses in Partial Split
Supersymmetry}
\author{Marco Aurelio D\'\i az}
\author{Sebasti\'an Garc\'ia S\'aenz}
\author{Benjamin Koch}
\affiliation{
{\small Departamento de F\'\i sica, Pontificia Universidad Cat\'olica 
de Chile, Avenida Vicu\~na Mackenna 4860, Santiago, Chile 
}}
\begin{abstract}
Partial Split Supersymmetry with bilinear R-parity violation
allows to reproduce all neutrino mass and mixing parameters. 
The viable dark matter candidate in this model is the gravitino.
We study the hypothesis that both possibilities are true:
Partial Split Supersymmetry explains neutrino physics and
that dark matter is actually composed of gravitinos.
Since the gravitino has
a small but non-zero decay probability, its decay products could
be observed in astrophysical experiments.
Combining bounds from astrophysical photon spectra
with the bounds coming from the mass matrix in the neutrino sector
we derive a stringent upper limit for the allowed gravitino mass.
This mass limit is in good agreement with
the results of direct dark matter searches.
\end{abstract}
\date{\today}
\pacs{14.60.Pq, 12.60.Jv, 95.35.+d}

\maketitle
\section{Introduction}

Split supersymmetry (SS) was originally proposed to address some of the most 
conspicuous 
problems of supersymmetric models, which are fast proton decay and excessive 
flavor changing neutral currents and CP violation \cite{ArkaniHamed:2004fb}. In
SS the 
solution to these problems is 
accomplished by considering all squarks and sleptons very massive, with a mass scale 
$\widetilde m$ somewhere between the supersymmetric scale $M_{susy}$ and the Grand 
Unification scale $M_{GUT}$. One of the Higgs bosons remains light, as usual in 
supersymmetric models, as well as the gauginos and higgsinos, with all these particles 
having a mass accessible to the LHC \cite{Aad:2009wy}.

If R-Parity violation (RpV) is introduced in supersymmetric models, lepton number 
and/or baryon 
number will be violated as well, inducing a potentially too fast proton 
decay \cite{Barbier:2004ez}.
Nevertheless, in SS the trilinear RpV couplings play little role in the phenomenology, 
with the exception being in the gluino decay rate. In this case, only bilinear RpV (BRpV) is 
relevant, 
opening up the possibility for a neutrino mass generation mechanism, 
without running into the danger
of a too fast proton decay. As in any BRpV model,
in SS-BRpV the atmospheric neutrino mass 
squared difference is generated by a low energy see-saw mechanism due to the mixing 
between neutrinos and neutralinos \cite{Nowakowski:1995dx,Hirsch:2000ef}.
Nevertheless, 
at one loop the only contributions to 
the neutrino mass matrix, coming from loops with the Higgs boson and neutralinos, is 
not enough to generate a solar neutrino mass squared 
difference \cite{Chun:2004mu,Diaz:2006ee}. Thus, an additional 
contribution to the model is needed \cite{Diaz:2009yz}
or the model itself has to be generalized.

In Partial Split Supersymmetry (PSS) all squarks and sleptons have a mass of the 
order of
the SS mass scale $\widetilde m$, but both Higgs doublets remain with a mass at the 
electroweak scale \cite{Sundrum:2009gv,Diaz:2006ee}. The addition of RpV to this
model was introduced to be able to generate a solar neutrino 
mass \cite{Diaz:2006ee}. Loop contributions from neutral CP-even and CP-odd 
Higgs bosons are indeed able to do the job, producing not only the atmospheric and 
solar masses, but also the atmospheric, solar, and reactor neutrino mixing 
angles \cite{Diaz:2009gf}.

Since R-Parity is not conserved, the lightest supersymmetric particle (LSP) will be unstable
and decay into SM particles. This will occur very fast if the LSP is the traditional
neutralino, losing it as a candidate for Dark Matter. It is however known
that in the case of a gravitino LSP, despite being unstable, it will decay very
slowly via
gravitationally induced couplings \cite{Grefe:2008zz,Wang:2005at}. 
Suppression of the gravitino 
interactions by both 
the Planck mass and the R-parity violating couplings leads to a very long-lived 
massive 
particle, whose lifetime can typically be of several orders of magnitude longer 
than the 
age of the universe. The next-to-lightest superparticle (NLSP), on the other hand, 
has a lifetime which is much shorter than 1 second, and thus becomes harmless to a 
successful big-bang nucleosynthesis \cite{Takayama:2000uz}.

The fact that the gravitino decays to ordinary particles in such
a scenario has given 
rise to interesting phenomenological studies 
\cite{Ibarra:2007wg,Ishiwata:2008cu,Yuksel:2007dr}. An important one is the study of 
the photon spectrum produced from the two-body decay 
\cite{Choi:2009ng,Buchmuller:2007ui,Bertone:2007aw}, and more recently from the 
three-body decays of the gravitino \cite{Choi:2010jt,Choi:2010xn}. Important
constraints on the 
mass and lifetime of the gravitino can be derived from the fact that its decay has not 
been detected by gamma-ray telescopes, most importantly the 
Fermi Large Area Telescope (Fermi LAT)
\cite{Collaboration:2010gq,Abdo:2010nz,Atwood:2009ez,Abdo:2009mr,Abdo:2010nc}.

Further 
constraints on the allowed parameters 
can be derived  from the measurements of the
neutrino 
masses and mixing parameters, which in PSS provide constraints on the R-parity 
violating couplings \cite{Diaz:2006ee,Diaz:2009gf}. 

In this work we study these two independent constraints, and show 
that when taken together they imply the existence of a maximal value of the 
light gravitino mass $m_{3/2}^{max}$, and also the existence of a minimum value of
the gravitino lifetime.
Our findings are in good agreement with direct dark matter
searches \cite{Ahmed:2009zw,Aprile:2011hi}
which put much stronger constraints on dark matter particles
with masses $\gg m_{3/2}^{max}$.
It is important to mention that this kind of model
can be further studied in the context of the observed Pamela
electron/positron excess and direct LHC signals
\cite{Bajc:2010qj,Chen:2009ew}.
Further studies are possible in the context of 
the early universe \cite{Khlopov:1984pf,Falomkin:1984eu,Khlopov:2004tn} .
However, since in our work the gravitino has an extremely long
lifetime, the implications from gravitino decay in the early universe
are not taken into account.

This model is a bottom up approach, it can however be motivated
by a number of general considerations. In 
ref.~\cite{ArkaniHamed:2004yi} it is shown that a Split Supersymmetric
spectrum can be easily generated in models with direct mediation of supersymmetry
breaking, {\sl i.e.}, without invoking a hidden sector. In opposition to low energy
supersymmetric models, in SS this is possible because the gauginos are 
allowed to be much lighter than the sfermions. A toy model is
given, where sfermion masses are generated at a scale $\widetilde m$ while 
gaugino masses are induced at a lower scale $\mu$, after integrating out 
physics at a higher scale that controls the supersymmetry breaking.
In the same model, a gravitino mass is generated at the scale 
$\widetilde m^2/M_{Pl}$ allowing it to be even lighter than gauginos.
This toy model generates also a $B_\mu$ term of the order of $\widetilde m^2$
and it would have to be modified in order to accommodate PSS, where $B_\mu$
is required to be much smaller than $\widetilde m^2$. It is worth to mention
that a $B_\mu\ll\widetilde m^2$ would alleviate the otherwise present 
fine-tuning needed in the Higgs potential to generate a correct 
electroweak symmetry braking \cite{Drees:2005cp}. 
As a top down approach to PSS we mention also
ref.~\cite{Sundrum:2009gv} where a PSS spectrum is given. In this model
sfermions masses are generated at a scale $\widetilde m^2\sim V_0/M^2$,
where supersymmetry breaking originates with a hidden sector dynamics
with vacuum energy $V_0$, and it is communicated to the visible sector by 
massive fields of mass $M$. In addition, two light Higgs scalars are
composite with a mass $m_h\sim\frac{g}{4\pi}\Lambda_{comp}$, with
$\Lambda_{comp}$ the typical scale of the strong dynamics. The gravitino
mass is of the order $\sqrt{V_0}/M_{Pl}$ and can be the LSP for values
for example $V_0\sim(10^9\,{\mathrm{GeV}})^4$.

\section{Partial Split Supersymmetry}

In PSS both Higgs doublets remain with a mass at the electroweak scale. As
it happens in SS, higgsinos, gauginos, and Higgs bosons interact via induced 
couplings of the type,
\begin{eqnarray}
{\cal L}_{PSS}^{RpC} &\owns& 
-\textstyle{\frac{1}{\sqrt{2}}} H_u^\dagger
(\tilde g_u \sigma\widetilde W + \tilde g'_u\widetilde B)\widetilde H_u
-\textstyle{\frac{1}{\sqrt{2}}} H_d^\dagger
(\tilde g_d \sigma \widetilde W - \tilde g'_d \widetilde B)\widetilde H_d 
+\mathrm{h.c.},
\label{RpCterms}
\end{eqnarray}
where $\tilde g_u$, $\tilde g'_u$, $\tilde g_d$, and $\tilde g'_d$ are
couplings induced in the effective low energy lagrangian. At the SS scale
$\widetilde m$ they satisfy the boundary conditions,
\begin{equation}
\tilde g_u=\tilde g_d=g\,,\qquad \tilde g'_u=\tilde g'_d=g',
\label{gMatching} 
\end{equation}
evolving with independent RGE down to the electroweak scale. Similarly to the
MSSM, both Higgs fields acquire a vacuum expectation value $\langle H_u\rangle=v_u$
and $\langle H_d\rangle=v_d$, with the constraint 
$v^2=v_u^2+v_d^2=246\,{\mathrm{GeV}}^2$
and the definition $\tan\beta=v_u/v_d$. Gauginos and higgsinos mix forming the
neutralinos, with a mass matrix very similar to the one in the MSSM,
\begin{equation}
{M}_{\chi^0}=\left[\begin{array}{cccc}
M_1 & 0 & -\frac{1}{2}\tilde g'_dc_\beta v & \frac{1}{2}\tilde g'_us_\beta v \\
0 & M_2 & \frac{1}{2}\tilde g_dc_\beta v & -\frac{1}{2}\tilde g_us_\beta v \\
-\frac{1}{2}\tilde g'_dc_\beta v & \frac{1}{2}\tilde g_dc_\beta v & 0 & -\mu \\
\frac{1}{2}\tilde g'_us_\beta v & -\frac{1}{2}\tilde g_us_\beta v & -\mu & 0
\end{array}\right],
\label{X0massmat2}
\end{equation}
where $M_1$ and $M_2$ are the gaugino masses associated to the $U(1)$ and
$SU(2)$ gauge bosons, and $\mu$ is the higgsino mass. In our calculations we will
neglect the running of the higgsino-higgs-gaugino couplings and work with 
their approximated value indicated by eq.~(\ref{gMatching}).

The addition of R-Parity violation to the PSS lagrangian allows us to study a
mechanism for neutrino mass generation. Trilinear couplings are not relevant for 
this problem, because all squarks and sleptons are very heavy, with a mass of the 
order of $\widetilde m$, and thus decoupled from the low energy effective
theory. Only BRpV is relevant, and is described by the terms,
\begin{equation}
{\cal L}_{PSS}^{RpV} =
-\epsilon_i \widetilde H_u^T \epsilon L_i  
\ -\ 
\textstyle{\frac{1}{\sqrt{2}}} b_i H_u^T\epsilon
(\tilde g_d \sigma\widetilde W-\tilde g'_d\widetilde B)L_i 
\ + \ h.c., 
\label{LSS2HDMRpV}
\end{equation}
where $\epsilon=i\sigma_2$. The $\epsilon_i$ are the supersymmetric BRpV
parameters in the superpotential, which at the low scale manifest themselves as 
mixing between higgsinos and lepton fields. The $b_i$ are three dimensionless 
parameters attached to lepton-higgs-gaugino interactions. They are analogous 
to the ones in eq.~(\ref{RpCterms}), except that violate R-Parity, and are
generated in the effective low energy theory.

The  origin of the BRpV terms in (\ref{LSS2HDMRpV}) is related to the
$\mu$-problem which refers to the origin of the term $\mu H_uH_d$ in the 
superpotential. As discussed for example in ref.~\cite{Allanach:2003eb}, the 
same mechanism that solve the $\mu$-problem could be used to explain the 
origin of the $\epsilon$ terms. A popular mechanism is the existence of
non renormalizable couplings of the sort $(a/M_{Pl})\Phi_1\Phi_2 H_u H_d$,
with $a$ being a dimensional constant and $\Phi_1$ and $\Phi_2$ two hidden sector 
scalars. When supersymmetry breaks and $\Phi_1$ and $\Phi_2$ acquire vacuum 
expectation values a $\mu$-term is generated. Similar terms 
$(a_i/M_{Pl})\Phi_3\Phi_4 H_u L_i$ can be present generating $\epsilon$-terms,
although $a_i$ and/or $\langle\Phi_3\rangle$, $\langle\Phi_4\rangle$
should be much smaller than $a$ and/or $\langle\Phi_1\rangle$, $\langle\Phi_2\rangle$
in order to have $\epsilon_i$ a few order of magnitude smaller than $\mu$,
necessary for neutrino physics. The $b_i$ terms are generated after integrating 
out the sleptons, and appear because above the scale $\widetilde m$ the
Higgs bosons mix with sleptons. At the scale $\widetilde m$ we have $b_i\sim v_i/v_u$,
with the necessary condition $v_uB_{\epsilon i}\sim v_i M_{Li}^2$
from the minimization of the scalar potential \cite{Diaz:2006ee}. Large
$B_{\epsilon i}$ can be easily obtained in a similar way as for $B_\mu$,
as explained in \cite{ArkaniHamed:2004yi}. We mention also 
ref.~\cite{FileviezPerez:2010ek} where 
R-Parity is naturaly broken radiatively when a right-handed 
sneutrino acquires a vacuum expectation value, generating bilinear R-Parity
violating terms. In our work, we concentrate on the effect of bilinear
R-Parity violation. Trilinear R-Parity violating couplings could be present,
but we ignore their effects due to the large mass of the sfermions.

\section{Neutrino masses}
\label{secNumass}

When Higgs bosons acquire vacuum expectation values, mixing terms between
gauginos and leptons are generated, producing the following mixing
between neutrinos and heavier fermions,
\begin{equation}
{\cal L}^{RpV}_{PSS} = - \left[
\epsilon_i \widetilde H_u^0 + \frac{1}{2} b_i v_u \left( 
\tilde g_d \widetilde W_3 - \tilde g'_d \widetilde B \right)
\right] \nu_i \ + \ h.c. \ + \ \ldots
\label{XnuMixing}
\end{equation}
The effect of these terms is that neutralinos mix with neutrinos, forming a 
$7\times7$ mass matrix, which in the basis 
$\left(\widetilde{B},\widetilde{W},\widetilde{H}_u,\widetilde{H}_d,
\nu_e,\nu_{\mu},\nu_{\tau}\right)$ has the form,
\begin{equation}
{\cal M}_N=\left[\begin{array}{cc} M_{\chi^0} 
& m^T \\ 
m & 0 \end{array}\right],
\label{X07x7}
\end{equation}
and where the $4 \times 4$ submatrix $M_{\chi^0}$ is the neutralino mass matrix
in eq.~(\ref{X0massmat2}). The $4\times3$ neutralino-neutrino mixing block
\begin{equation}
m=\left[\begin{array}{cccc}
-\frac{1}{2} \tilde g'_d b_1 v_u & 
 \frac{1}{2} \tilde g_d  b_1 v_u & 
0 &\epsilon_1 
\cr
-\frac{1}{2} \tilde g'_d b_2 v_u & 
 \frac{1}{2} \tilde g_d  b_2 v_u & 
0 & \epsilon_2 
\cr
-\frac{1}{2} \tilde g'_d b_3 v_u & 
 \frac{1}{2} \tilde g_d  b_3 v_u &
0 & \epsilon_3
\end{array}\right],
\end{equation}
develops from terms in eq.~(\ref{XnuMixing}). A diagonalization by blocks of the mass
matrix in eq.~(\ref{X07x7}) can be achieved by a rotation 
$\mathcal{N}\mathcal{M}_N\mathcal{N}^T$ given
by,
\begin{equation}
\mathcal{N}\simeq\begin{bmatrix} N & N\xi^T \\ -\xi & 1 \end{bmatrix},
\label{Nmatrix}
\end{equation}
where we define,
\begin{equation}
\xi=m M_{\chi^0}^{-1},
\end{equation}
with
\begin{eqnarray}
\xi_{i1}&=&\frac{\tilde g'_d \, \mu M_2}{2\det{M_{\chi^0}}}\Lambda_i
\nonumber\\
\xi_{i2}&=&-\frac{\tilde g_d \, \mu M_1}{2\det{M_{\chi^0}}}\Lambda_i
\label{xis}\\
\xi_{i3}&=&\frac{v_u}{4\det{M_{\chi^0}}}\left(
M_1\tilde g_u \tilde g_d + M_2\tilde g'_u\tilde g'_d 
\right)\Lambda_i - \frac{\epsilon_i}{\mu}
\nonumber\\
\xi_{i4}&=&-\frac{v_d}{4\det{M_{\chi^0}}}\left(
M_1\tilde g^2_d + M_2\tilde g'^2_d\right)\Lambda_i
\nonumber
\end{eqnarray}
and $\Lambda_i=\mu b_i v_u + \epsilon_i v_d$. This leaves an induced effective 
neutrino mass matrix equal to,
\begin{equation}
{\bf M}_\nu^{(0)}|_{ij}=
-m^T {\mathrm{M}}_{\chi^0}^{-1} m|_{ij}=
A^{(0)} \Lambda_i\Lambda_j,
\label{treenumass2}
\end{equation}
with
\begin{equation}
A^{(0)}= \frac{M_1 \tilde g^2_d + M_2 \tilde g'^2_d}{4\det{M_{\chi^0}}}.
\label{A0}
\end{equation}
At this level only one neutrino
acquires mass, leaving the solar squared mass difference null and the solar angle
undetermined. Quantum corrections contribute to the neutrino mass matrix in such 
a way that the degeneracy in eq.~(\ref{treenumass2}) is lifted, leaving it with 
the following form,
\begin{equation}
{\bf M}_\nu|_{ij}=A\Lambda_i\Lambda_j+
C\epsilon_i\epsilon_j,
\label{DpiH2HDM}
\end{equation}
where the $B$ term \cite{Diaz:2009gf} has been made to vanish by an appropriate 
choice of the subtraction point. Thus at one loop two neutrinos acquire a mass
while the third one remains massless. Since in this case the experimental value
$\Delta m^2_{sol}/\Delta m^2_{atm}\approx 0.035$ implies $m_{\nu_3}\gg m_{\nu_2}$
we have,
\begin{eqnarray}
\Delta m^2_{atm} &\approx& \left( A|\vec\Lambda|^2+C|\vec\epsilon\,|^2 \right)^2 -
2AC|\vec\Lambda\times\vec\epsilon|^2,
\nonumber\\
\Delta m^2_{sol} &\approx& \frac{A^2C^2|\vec\Lambda\times\vec\epsilon\,|^4}
{\left( A|\vec\Lambda|^2+C|\vec\epsilon|^2 \right)^2}.
\label{numass2}
\end{eqnarray}

For later use, we introduce the photino $\widetilde\gamma$ and zino $\widetilde Z$ 
fields by rotating by the weak mixing angle the weakly interacting gauginos
$\tilde B, \tilde W$, in direct analogy to their standard model counter parts
\begin{equation}
 \left(
\begin{array}{c}
\widetilde \gamma\\
\widetilde Z^0 \\
\dots
\end{array}
\right)=
\left(
\begin{array}{ccc}
 c_W & s_W \dots \\
-s_W & c_W \dots \\
\dots&\dots
\end{array}
\right)
\left(
\begin{array}{c}
\widetilde B\\
\widetilde W\\
\dots
\end{array}
\right)=
A_W
\left(
\begin{array}{c}
\widetilde B\\
\widetilde W\\
\dots
\end{array}
\right),
\end{equation}
where the dots indicate that all other states
are just multiplied by the unit matrix.
Thus when dealing with this new basis
$\left(\widetilde{\gamma},\widetilde{Z^0},\widetilde{H}_u,\widetilde{H}_d,
\nu_e,\nu_{\mu},\nu_{\tau}\right)$
the mixing matrix is
\begin{equation}\label{NwithThetaW}
 \mathcal{N}'=\mathcal{N} A_W^T\quad,
\end{equation}
where only the first two states are rotated.

\section{Gravitino Decay}

Indirect observation of the gravitino becomes a possibility due to its decay to 
ordinary particles. In this section we calculate the possible decay channels of the 
gravitino as dark matter candidate, assuming that $m_{3/2}<m_W$. We then relate these 
results to experimental bounds on the decay products and on the RpV parameters.

\subsection{Two-body decay}

When R-parity is conserved the gravitino can radiatively decay into a photon via 
the following term in the Lagrangian,
\begin{equation} \label{coupleGravi}
{\cal L} \owns -\frac{1}{4M_P} \overline\psi_\mu \sigma^{\nu\rho}
\gamma^\mu \lambda_\gamma F_{\nu\rho}
\end{equation}
where $M_P$ is the Planck mass, $\psi_\mu$ is the spin-$3/2$ gravitino 
field, $\lambda_\gamma$ is the spin-$1/2$ photino field, 
$F_{\nu\rho}=\partial_\nu A_\rho-\partial_\rho A_\nu$ is the photon field strength, 
and $A_\mu$ is the photon field. This coupling might in principle be modified by a 
factor of order one. In this work however 
we assume that (\ref{coupleGravi}) is the exact form of the coupling.
Variations of the final result can then be studied
by order one shifts of the Planck mass $M_P$. This Lagrangian term induces the following decay,
\begin{center}
\vspace{-50pt} \hfill \\
\begin{picture}(110,100)(10,45) 
\SetWidth{0.7}
\ArrowLine(0,50)(50,50)
\Photon(0,50)(50,50){5}{4}
\Photon(50,50)(100,100){5}{4.5}
\ArrowLine(50,50)(75,25)
\ArrowLine(75,25)(100,0)
\Line(72,25)(78,25)
\Line(75,28)(75,22)
\Text(0,58)[lb]{$\widetilde{G}$}
\Text(105,90)[lb]{$\gamma$}
\Text(55,20)[lb]{$\widetilde{\gamma}$}
\Text(105,0)[lb]{$\nu_i$}
\end{picture}
$
=-\,i\,{\cal M}_{0}
$
\vspace{30pt} \hfill \\
\end{center}
\vspace{40pt}
where the cross indicates we are picking the photino component of the corresponding
neutrino, which mix due to violation of R-Parity. The amplitude for this decay is
\begin{equation}
{\cal M}_0=-\frac{i}{4M_P}\Big\{ \overline\nu(q,s) \gamma^{\mu} 
\big[\slash\!\!\! k,\gamma^{\nu}\big]
\psi_{\mu}(p,\lambda) 
\Big\} \epsilon_{\mu}(k,m) \, {U}_{\widetilde{\gamma}\nu},
\end{equation}
where ${U}_{\widetilde\gamma\nu}$ is the amount of photino in the neutrino fields, as 
indicated by the neutrino eigenvector. We write the gravitino field as the tensor 
product of a spin-1/2 field with a spin-1 field,
\begin{equation}
\psi_{\mu}(p,\lambda)=\sum_{s,m} 
\langle 1/2,s;1,m|3/2,\lambda \rangle u(p,s) \epsilon_\mu(p,m),
\end{equation}
obtaining the following completeness relation \cite{Grefe:2008zz},
\begin{equation}
\sum_\lambda \psi_\mu(p,\lambda)\overline\psi_\nu(p,\lambda) =
-(\slash\!\!\! p-m_{3/2}) \Bigg[ 
\bigg( g_{\mu\nu}-\frac{p_\mu p_\nu}{m_{3/2}^2} \bigg)
-\frac{1}{3} 
\bigg( g_{\mu\sigma}-\frac{p_\mu p_\sigma}{m_{3/2}^2} \bigg)
\bigg( g_{\nu\lambda}-\frac{p_\nu p_\lambda}{m_{3/2}^2} \bigg)
\gamma^\sigma \gamma^\lambda \Bigg].
\end{equation}
Knowing the above relation, the calculation of the differential cross section is 
standard, giving the result
\begin{equation}
\frac{d\Gamma}{dE_\gamma d\Omega}=
\frac{\langle|{\cal {M}}_0|^2\rangle}
{64\pi^2m_{3/2}}\delta\left(E_\gamma-\frac{m_{3/2}}{2}\right)
=\frac{m_{3/2}^3}{128\pi^2M_P^2}|{U}_{\widetilde\gamma\nu}|^2 
\delta\left(E_\gamma-\frac{m_{3/2}}{2}\right),
\end{equation}
which is independent of the angles as expected. The total decay rate is then
\begin{equation} \label{eq:2body}
\Gamma(\widetilde{G}\rightarrow\gamma\nu) = 
\frac{m_{3/2}^3}{32\pi M_P^2}|{U}_{\widetilde{\gamma}\nu}|^2.
\end{equation}
For a gravitino mass $m_{3/2}<m_W$ this is the only kinematically allowed 2-body 
decay. By using the relation in eq.~(\ref{NwithThetaW})
one finds that the photino-neutrino mixing
factor in PSS is
\begin{equation}
U_{\widetilde{\gamma}\nu_i}=\mathcal{N}_{i1} c_W+\mathcal{N}_{i2} s_W,
\end{equation}
with $i=5,6,7$ labeling the neutrino generation, and where $t_W=g'/g$ is the tangent 
of the weak mixing angle. Using equations (\ref{Nmatrix}) to (\ref{xis}) we find,
\begin{equation} \label{eq:gammamixing}
U_{\widetilde{\gamma}\nu_i}\simeq\frac{\mu}{2({\rm det}M_{\chi^0})} 
\left(\tilde{g}_dM_1s_W-\tilde{g}'_dM_2c_W\right)\Lambda_i.
\end{equation}
For the numerical calculations we will require the sum over the generations of the 
square of the mixing factor:
\begin{equation} \label{eq:gammamixingsq}
|U_{\widetilde{\gamma}\nu}|^2:=\sum_{i} U_{\widetilde{\gamma}\nu_i}^2 \simeq 
\frac{\mu^2}{4({\rm det}M_{\chi^0})^2} 
\left(\tilde{g}_dM_1s_W-\tilde{g}'_dM_2c_W\right)^2|\vec{\Lambda}|^2.
\end{equation}
This is because we do not distinguish the different neutrino flavors.

\subsection{Three-body decays}

When studying the three body decay a more general part
of the interaction Lagrangian comes into play
\bea\label{coupleGravi2}
{\mathcal{L}}&\owns&
-\frac{i}{\sqrt{2}M_P}\left[
(D^*_\mu\phi^{i*})\bar \psi_\nu \gamma^\mu \gamma^\nu P_L \chi^i-
(D_\mu \phi^i)\bar \chi^i P_R \gamma^\nu \gamma^\mu \psi_\nu
\right]\\ \nonumber
&&-\frac{i}{8 M_P}\bar \psi_\mu 
\left[ \gamma^\nu,\gamma^\rho\right]
\gamma^\mu \lambda^{(\alpha)a}F_{\nu \rho}^{(\alpha)a}\quad,
\eea
where the second line is in analogy to (\ref{coupleGravi}) and the first line
introduces additional couplings with scalar fields $\phi$.
The 3-body decays of the gravitino were studied in detail for the first time in 
\cite{Choi:2010jt,Choi:2010xn}, where explicit formulae are given. Nevertheless,
our calculations 
have yielded that the three-body results in \cite{Choi:2010jt} have to be corrected.
We agree, however, with the conclusion that the 3-body decays are indeed important, 
and cannot be neglected. We find 3-body decay branching ratios of the order of $10\%$
for gravitino masses of order 10 GeV, and greater for larger masses. The exact 
formulas for the amplitudes of the contributing diagrams are given in the Appendix.

First we consider the gravitino decay into a fermion pair and a neutrino. The 
first pair 
of diagrams are three-body decays via an intermediate photon and $Z$ boson,
\begin{center}
\vspace{-50pt} \hfill \\
\begin{picture}(130,120)(40,45) 
\SetWidth{0.7}
\ArrowLine(0,50)(50,50)
\Photon(0,50)(50,50){5}{4}
\Photon(50,50)(85,85){5}{3.5}
\ArrowLine(85,85)(120,120)
\ArrowLine(120,50)(85,85)
\ArrowLine(50,50)(85,15)
\ArrowLine(85,15)(120,-20)
\Line(82,15)(88,15)
\Line(85,18)(85,12)
\Text(0,58)[lb]{$\widetilde{G}$}
\Text(48,70)[lb]{$\gamma,Z$}
\Text(125,110)[lb]{$f$}
\Text(125,50)[lb]{$\bar{f}$}
\Text(48,15)[lb]{$\widetilde{\gamma},\widetilde{Z}$}
\Text(125,-20)[lb]{$\nu_i$}
\end{picture}
$
=-\,i\,{\cal M}_{1,2}
$
\vspace{30pt} \hfill \\
\end{center}
\vspace{40pt}
where the cross means we take the photino (or zino) component in the neutrino field.
These two amplitudes are equal to,
\begin{eqnarray}
{\cal M}_1 &=& i
\frac{-i}{k^2} \overline{u}(k_1) (-iq_f) \gamma_\mu v(k_2)
\overline{u}(q)\frac{-iU_{\widetilde\gamma \nu}}{4M_P} \gamma^\alpha
( k\!\!\!/ \gamma^\mu - \gamma^\mu k\!\!\!/ ) \psi_\alpha(p),
\\
{\cal M}_2 &=& i
\frac{-i}{(k^2-m_Z^2)+im_Z\Gamma_Z}
\overline{u}(k_1) \frac{-ig}{c_W} \gamma_\mu(c_V^f+c_A^f \gamma_5)v(k_2)
\overline{u}(q)\frac{-iU_{\widetilde Z \nu}}{4M_P} \gamma^\alpha
( k\!\!\!/ \gamma^\mu - \gamma^\mu k\!\!\!/ ) \psi_\alpha(p),
\nonumber
\end{eqnarray}
with $c_V^f=T_3^f/2-q_fs_W^2$ and $c_A^f=-T_3^f/2$. Their contribution to the 
decay rate 
is given in the Appendix. The photino-neutrino
mixing is given in eq.~(\ref{eq:gammamixing}), while an analogous calculation for the
zino-neutrino mixing gives,
\begin{equation} \label{eq:zmixing}
U_{\widetilde{Z}\nu_i}\simeq-\frac{\mu}{2({\rm det}M_{\chi^0})} 
\left(\tilde{g}_dM_1c_W+\tilde{g}'_dM_2s_W\right)\Lambda_i.
\end{equation}
To the previous two amplitudes we add a contribution coming from quartic
couplings
between the gravitino, a gauge boson, and a scalar with its fermionic partner, with the
scalar acquiring a vacuum expectation value,
\begin{center}
\vspace{-50pt} \hfill \\
\begin{picture}(320,120)(10,45) 
\SetWidth{0.7}
\ArrowLine(0,50)(50,50)
\Photon(0,50)(50,50){5}{4}
\Photon(50,50)(85,85){5}{3.5}
\ArrowLine(85,85)(120,120)
\ArrowLine(120,50)(85,85)
\ArrowLine(50,50)(120,-20)
\DashLine(40,10)(50,50){5}
\Text(0,58)[lb]{$\widetilde{G}$}
\Text(50,70)[lb]{$Z$}
\Text(125,110)[lb]{$f$}
\Text(125,50)[lb]{$\bar{f}$}
\Text(125,-20)[lb]{$\nu_i$}
\Text(32,-4)[lb]{$\langle\tilde{\nu}_i\rangle$}
\Text(150,50)[lb]{$+$}
\ArrowLine(180,50)(230,50)
\Photon(180,50)(230,50){5}{4}
\Photon(230,50)(265,85){5}{3.5}
\ArrowLine(265,85)(300,120)
\ArrowLine(300,50)(265,85)
\ArrowLine(230,50)(265,15)
\ArrowLine(265,15)(300,-20)
\DashLine(220,10)(230,50){5}
\Line(262,15)(268,15)
\Line(265,18)(265,12)
\Text(180,58)[lb]{$\widetilde{G}$}
\Text(230,70)[lb]{$Z$}
\Text(305,110)[lb]{$f$}
\Text(305,50)[lb]{$\bar{f}$}
\Text(236,12)[lb]{$\widetilde{H_d}$}
\Text(305,-20)[lb]{$\nu_i$}
\Text(211,-4)[lb]{$\langle H_d \rangle$}
\end{picture}
$
=-\,i\,{\cal M}_{3}
$
\vspace{30pt} \hfill \\
\end{center}
\vspace{40pt}
The first amplitude is proportional to $b_i v_u$ and the second
to
$v_d(\epsilon_i/\mu)$,
in such a way that the combined amplitude ${\cal M}_{3}$ is proportional to 
$\Lambda_i$,
\begin{equation}
{\cal M}_3 = i
\frac{-i}{(k^2-m_Z^2)+im_Z\Gamma_Z}
\overline{u}(k_1) \frac{-ig}{c_W} \gamma_\mu(c_V^f+c_A^f \gamma_5)v(k_2)
\overline{u}(q) \frac{-i g\Lambda_i}{4c_W\mu M_P}
P_R\gamma^\alpha\gamma^\mu \psi_\alpha(p)\quad.
\label{M3}
\end{equation}
whose contribution to the decay rate is also given in the Appendix. Since the neutrino 
is not directly detected, a sum over flavors must be done [as in eq.~(\ref{eq:gammamixingsq})], after which it is clear that the 
decay rate will satisfy 
$\Gamma(\widetilde G\rightarrow f\overline{f} \nu)\propto|\vec\Lambda|^2$.

Now we consider the gravitino decay into two fermions and a charged lepton.
The decay via a $W$ gauge boson is represented by the Feynman diagram
\newpage
\begin{center}
\vspace{-50pt} \hfill \\
\begin{picture}(130,120)(40,45) 
\SetWidth{0.7}
\ArrowLine(0,50)(50,50)
\Photon(0,50)(50,50){5}{4}
\Photon(50,50)(85,85){5}{3.5}
\ArrowLine(85,85)(120,120)
\ArrowLine(120,50)(85,85)
\ArrowLine(50,50)(85,15)
\ArrowLine(85,15)(120,-20)
\Line(82,15)(88,15)
\Line(85,18)(85,12)
\Text(0,58)[lb]{$\widetilde{G}$}
\Text(50,70)[lb]{$W$}
\Text(125,110)[lb]{$f$}
\Text(125,50)[lb]{$\bar{f'}$}
\Text(52,16)[lb]{$\widetilde{W}$}
\Text(125,-20)[lb]{$\ell_i^-$}
\end{picture}
$
=-\,i\,{\cal M}_{4}
$
\vspace{30pt} \hfill \\
\end{center}
\vspace{40pt}
whose amplitude can be shown to be,
\begin{equation}
{\cal M}_4 = i
\frac{-i}{(k^2-m_W^2)+im_W\Gamma_W}
\overline{u}(k_1) \frac{-ig}{2\sqrt{2}} \gamma_\mu(1- \gamma_5)v(k_2)
\overline{u}(q)\frac{-iU_{\widetilde W \ell}}{4 M_P} \gamma^\alpha
( k\!\!\!/ \gamma^\mu - \gamma^\mu k\!\!\!/ ) \psi_\alpha(p).
\nonumber
\end{equation}
Its contribution to the decay rate is given in the Appendix, and it is proportional to
the wino mixing to charged leptons, given by,
\begin{equation} \label{eq:wmixing}
U_{\widetilde{W}\ell_i}\simeq-\frac{\tilde{g}_d}{\sqrt{2}
({\rm det}M_{\chi^{+}})}\Lambda_i,
\end{equation}
where $M_{\chi^{+}}$ is the chargino mass matrix \cite{Diaz:2006ee}. This graph is 
complemented by an amplitude coming from quartic couplings between the gravitino, a 
$W$ gauge boson, and a neutral scalar with its charged fermionic partner, with the
scalar acquiring a vacuum expectation value,
\begin{center}
\vspace{-50pt} \hfill \\
\begin{picture}(320,120)(10,45) 
\SetWidth{0.7}
\ArrowLine(0,50)(50,50)
\Photon(0,50)(50,50){5}{4}
\Photon(50,50)(85,85){5}{3.5}
\ArrowLine(85,85)(120,120)
\ArrowLine(120,50)(85,85)
\ArrowLine(50,50)(120,-20)
\DashLine(40,10)(50,50){5}
\Text(0,58)[lb]{$\widetilde{G}$}
\Text(50,70)[lb]{$W$}
\Text(125,110)[lb]{$f$}
\Text(125,50)[lb]{$\bar{f'}$}
\Text(125,-20)[lb]{$\ell_i^-$}
\Text(32,-4)[lb]{$\langle\tilde{\nu}_i\rangle$}
\Text(150,50)[lb]{$+$}
\ArrowLine(180,50)(230,50)
\Photon(180,50)(230,50){5}{4}
\Photon(230,50)(265,85){5}{3.5}
\ArrowLine(265,85)(300,120)
\ArrowLine(300,50)(265,85)
\ArrowLine(230,50)(265,15)
\ArrowLine(265,15)(300,-20)
\DashLine(220,10)(230,50){5}
\Line(262,15)(268,15)
\Line(265,18)(265,12)
\Text(180,58)[lb]{$\widetilde{G}$}
\Text(230,70)[lb]{$W$}
\Text(305,110)[lb]{$f$}
\Text(305,50)[lb]{$\bar{f'}$}
\Text(236,12)[lb]{$\widetilde{H}_d^-$}
\Text(305,-20)[lb]{$\ell_i^-$}
\Text(211,-4)[lb]{$\langle H_d \rangle$}
\end{picture}
$
=-\,i\,{\cal M}_{5}
$
\vspace{30pt} \hfill \\
\end{center}
\vspace{40pt}
As before, the first amplitude is proportional to $b_i v_u$ and the second to 
$v_d(\epsilon_i/\mu)$, such that the combined amplitude ${\cal M}_{5}$ is 
proportional to $\Lambda_i$,
\begin{equation}
{\cal M}_5 = i
\frac{-i}{(k^2-m_W^2)+im_W\Gamma_W}
\overline{u}(k_1) \frac{-ig}{2\sqrt{2}} \gamma_\mu(1- \gamma_5)v(k_2)
\overline{u}(q) \frac{-ig\Lambda_i}{2\sqrt{2} \mu M_P}
P_R\gamma^\alpha\gamma^\mu \psi_\alpha(p).
\label{M5}
\end{equation}
Its contribution to the decay rate is also given in the Appendix. Note that when 
fermion $f$ is a charged lepton and $f'$ is a neutrino, we get interference between 
diagrams ${\cal M}_{1,2,3}$ in one hand and ${\cal M}_{4,5}$ in the other. This is 
because the decay $\widetilde G\rightarrow \ell^+_i\ell^-_i\nu_j$ can proceed via a 
$Z$ or a $W$ gauge boson. In addition, note that 
$\Gamma(\widetilde G\rightarrow f\overline{f}'\ell^-_i)=
\Gamma(\widetilde G\rightarrow f'\overline{f}\ell^+_i)$, thus we multiply the first one 
by two.

\begin{figure}
\begin{center}
\includegraphics[angle=270,width=0.6\textwidth]{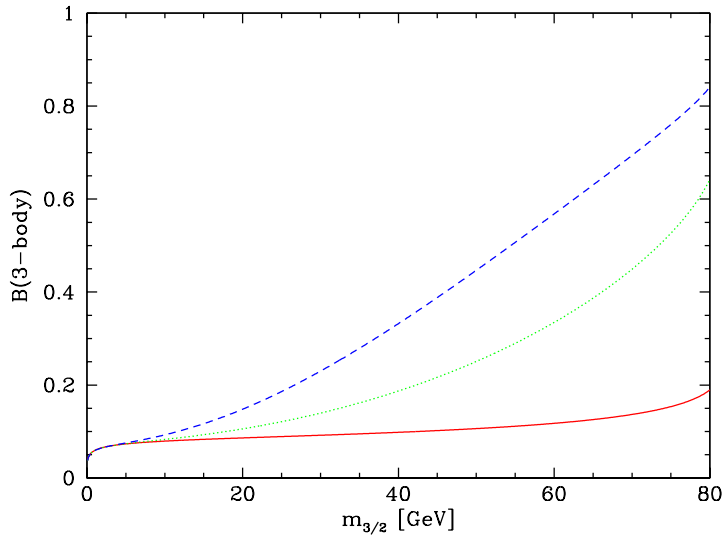}
\caption{\it Branching ratio of the three-body decay of the gravitino as a
function of its mass
$m_{3/2}$, red (full line) for $M_1=100\,$GeV, green (dotted line) for $M_1=300\,$GeV, 
and blue (dashed line) for $M_1=500\,$GeV.}
\label{bratio1p}
\end{center}
\end{figure}

The decay rate for the 3-body decays of a gravitino can be written as a sum of 
the various terms given in the Appendix. We sum over three generations, and 
neglect the masses of the final states. The end result is proportional to 
$|\vec{\Lambda}|^2$. Thus, in the branching ratio the factor $|\vec{\Lambda}|^2$ 
cancels, depending only on $M_1$, $M_2$, and the gravitino mass $m_{3/2}$
\cite{Choi:2010jt}. Here we work with the assumption $M_2\simeq 2M_1$. The result 
is shown in Fig.~\ref{bratio1p}, where we plot the branching ratio as a function of 
$m_{3/2}$ for the three values of $M_1=$ 100, 300, and 500 GeV. One observes
that the three-body decay becomes important for large
gravitino masses $m_{3/2}$ and large $M_1$.
%
%
The dependency on $m_{3/2}$ can be understood as a phase-space effect
in the three-body decay rate. The influence of $M_1$ on the branching ratio 
can be understood by the fact that the two-body and three-body decays, by the 
virtue of the mixings $U_{\tilde\gamma\nu}$ and $U_{\tilde Z\nu}$ 
respectively, get suppressed by growing $M_1$. In contrast the parts of the 
three-body decays that contain vacuum expectation values 
$\langle\tilde \nu_i\rangle$ and $\langle H_d \rangle$ do not experience 
this suppression and thus become more important in the regime of large 
$M_1$. Those effects are also present in \cite{Choi:2010jt}, it is only the 
form of the curves that turned out to change in the corrected version. For 
the masses we are interested here, $m_{3/2}\lsim10$ GeV, the 3-body decay 
is $\lsim10\%$ and with small dependence on the gaugino mass. For larger 
gravitino masses the branching ratio can be as large as $80\%$. Since the 
calculations are being carried out in the Feynman gauge, there are in 
principle also diagrams containing Goldstone bosons in the propagator. 
However, due to the coupling of the Goldstone bosons, those contributions 
vanish in the limit of light fermion masses.

\subsection{Induced photon flux}

The photon spectrum produced by the decay of the gravitino consists of a 
mono-energetic line of energy $m_{3/2}/2$ from the two-body decay, plus a 
continuum distribution from the three-body decays. The exact form of the 
spectrum, which depends on $m_{3/2}$ and $M_1$, was studied in detail in 
\cite{Choi:2010jt,Choi:2010xn} using an event generator. Here we are interested
in 
obtaining constraints on the gravitino parameters, for which it suffices 
as an approximation to consider only the photon line from the two-body 
decay, as this is the most prominent feature of the spectrum for values 
of $M_1$ up to 1 TeV \cite{Choi:2010jt}.
Including the three-body decay would make our final conclusions
slightly more stringent.

The observed spectrum is calculated from the flux of gamma-rays expected at 
earth. This flux is the sum of two contributions, one from the gravitinos 
decaying in the galactic halo, and one from the gravitinos decaying at 
cosmological distances. It has been shown that the first contribution is 
highly dominant, and so we will neglect the second one \cite{Buchmuller:2007ui}. 
In this way the differential flux, as a function of the photon energy $E$, 
has the following simple form \cite{Bertone:2007aw,Buchmuller:2007ui}:
\begin{equation} \label{eq:flux1}
E^2\frac{dJ_{\rm{halo}}}{dE}
=\frac{D_{\gamma}m_{3/2}}{2}\delta\left(E-\frac{m_{3/2}}{2}\right),
\end{equation}
with
\begin{equation}
D_{\gamma}=\frac{\Gamma(\widetilde{G}\rightarrow\gamma\nu)}{8\pi}
\left\langle\int_{\rm{l.o.s.}}\rho_{\rm{halo}
}(\vec{r})d\ell \right\rangle =
d_{\gamma}\Gamma(\widetilde{G}\rightarrow\gamma\nu).
\label{Dgamma}
\end{equation}
The constant $d_{\gamma}$ depends on the dark matter density profile of the halo, 
and on the region of the sky considered for averaging the flux (denoted by the 
term in brackets above, where {\sl l.o.s.} means {\sl line of sight}). Using the
Navarro-Frenk-White profile \cite{Navarro:1995iw}, and considering the region 
$|b|\geq10^{\circ}$ for the average (with $b$ denoting the latitude in galactic coordinates), we find
\begin{equation} \label{eq:dgamma}
d_{\gamma}=0.80\times10^{24}~~[{\rm{MeV~cm^{-2}~str^{-1}}}].
\end{equation}
This region of the sky was the one considered by the Fermi LAT collaboration 
in the derivation of the extragalactic diffuse spectrum \cite{Abdo:2010nz}.

\section{Limits and Constraints}

\subsection{Constraint from the observed photon spectrum}

The fact that the line produced by the gravitino two-body decay has not been observed 
gives constraints on the mass and the lifetime of the gravitino. Assuming that
the
extragalactic diffuse spectrum measured by Fermi LAT can be correctly modeled in 
terms of known sources \cite{Collaboration:2010gq}, one can use this spectrum to
find constraints  \cite{Abdo:2010nz,Buchmuller:2009xv}.

After the convolution between the calculated flux in eq.~(\ref{eq:flux1}) and a 
Gaussian distribution we find
\begin{equation}
E^2\frac{dJ_{\rm{halo}}}{dE}=\frac{D_{\gamma}m_{3/2}}{2}\frac{1}{\sqrt{
2\pi\sigma^2}}e^{-(E-m_{3/2}/2)^2/2\sigma^2},
\end{equation}
where $\sigma$ is related to the energy-dependent resolution of the Fermi LAT 
instrument
$p$ evaluated at the 2-body peak
\begin{equation}
\sigma=pE=p\frac{m_{3/2}}{2}\quad.
\end{equation}
The energy dependence of $p$ evaluated at $E=m_{3/2}/2$ can be approximated by
\cite{Atwood:2009ez,web}
\begin{equation}
p(m_{3/2}) = 0.349-0.142\log\left(\frac{m_{3/2}}{2~
{\rm MeV}}\right)+0.019\log^2\left(\frac{m_{3/2}}{2~{\rm MeV}}\right).
\end{equation}
Thus, the maximum of the photon spectrum is given by
\begin{equation}
\left(E^2\frac{dJ_{\rm{halo}}}{dE}\right)_{\rm{max}}=\frac{D_{\gamma}}{p(m_{3/2})\sqrt{
2\pi}}
\label{maxFlux}
\end{equation}
On the other hand, the intensity 
or the integrated flux for the extragalactic diffuse emission (for the range
$E>100$ MeV
and the sky region $|b|\geq10^{\circ}$) was measured by the Fermi LAT collaboration \cite{Abdo:2010nz},
\begin{equation} \label{eq:intensity}
I(>100~{\rm{MeV}})=\int_{100}^\infty\,\frac{dJ}{dE}=
(1.03\pm0.17)\times10^{-5}~[\rm{cm^{-2}~s^{-1}~str^{-1}}].
\end{equation}
together with the observation that the spectrum can be fitted by a power law 
$dJ/dE\propto E^{-\gamma}$, with index $\gamma=2.41\pm0.05$. 
From this 
we calculate the spectrum to be
\begin{equation} \label{eq:flux}
E^2\frac{dJ}{dE}=(9.6\pm1.6)\times10^{-3}~
[{\rm MeV~cm^{-2}~s^{-1}~str^{-1}}]~\left(\frac{E}{1~{\rm{MeV}}}\right)^{2-\gamma},
\end{equation}
where the error was directly calculated from the error in the integrated flux 
using eq.~(\ref{eq:intensity}).

As we mentioned before, we assume that the central value of the spectrum can be 
explained by models of known sources \cite{Collaboration:2010gq}. We impose that 
the extra contribution from the gravitino 
source is smaller than a $3\sigma$ error margin. This gravitino contribution 
is related to the decay rate $\Gamma(\widetilde{G}\rightarrow\gamma\nu)$ through
eqs.~(\ref{Dgamma}) and (\ref{maxFlux}). By introducing the total decay width, 
$\tau_{3/2}^{-1}=\Gamma(\widetilde{G}\rightarrow\gamma\nu)+
\Gamma(\mathrm{3-body})$, we find the following restriction on the gravitino 
lifetime,
\begin{equation} \label{eq:constraint1}
\left(\frac{\tau_{3/2}}{10^{27}~{\rm s}}\right)>\frac{0.851}{p(m_{3/2})}
B(\widetilde{G}\rightarrow\gamma\nu) 
\left(\frac{m_{3/2}}{1~{\rm GeV}}\right)^{\gamma-2},
\end{equation}
where $B(\widetilde{G}\rightarrow\gamma\nu)$ denotes the branching ratio of the 2-body decay. In this way eq.~(\ref{eq:constraint1}) defines a region in the 
$m_{3/2}-\tau_{3/2}$ plane consistent with the non-observation of the gravitino 
decay by Fermi LAT.

The gravitino 2-body decay width $\Gamma(\widetilde{G}\rightarrow\gamma\nu)$ is
given in eq.~(\ref{eq:2body}). If we sum over all neutrino species it becomes 
proportional to $|\vec\Lambda|^2$, via $|U_{\widetilde{\gamma}\nu}|^2$ in
eq.~(\ref{eq:gammamixingsq}). The 3-body decay width is also proportional to
$|\vec\Lambda|^2$. The reasons are analogous to the 2-body decay, since after
summing over lepton generations we see that $|{\cal M}_1|^2$ is proportional to 
$|U_{\widetilde{\gamma}\nu}|^2$, $|{\cal M}_2|^2$ is proportional to
$|U_{\widetilde{Z}\nu}|^2$ [eq.~(\ref{eq:zmixing})], and $|{\cal M}_4|^2$
is proportional to $|U_{\widetilde{W}\ell}|^2$ [eq.~(\ref{eq:wmixing})].
In addition, amplitudes $|{\cal M}_3|^2$ and $|{\cal M}_5|^2$ are directly
proportional to $|\vec\Lambda|^2$ as can be seen from eqs.~(\ref{M3}) and
(\ref{M5}). In this way, the gravitino lifetime becomes large for two reasons,
because the Planck mass is large and because BRpV is small: 
$\tau_{3/2}^{-1}\propto |\vec\Lambda|^2/M_P^2$.

We display experimental constraints on the model in the $m_{3/2}-\tau_{3/2}$ 
plane, with the first one given by eq.~(\ref{eq:constraint1}). This constraint
depends also on $|U_{\widetilde{\gamma}\nu}|^2$ and the gaugino masses. In fact,
the whole decay rate $\Gamma(\mathrm{3-body})$ can be factored out by
$|U_{\widetilde{\gamma}\nu}|^2$ with the remaining factors depending on $M_1$
and $M_2$, but with the dependence on $\mu$ being in first approximation 
negligible. We further use the simplifying assumption $M_2=2M_1$ motivated by
mSUGRA models. In this way, we display the constraints in the plane 
$m_{3/2}-\tau_{3/2}$ as a function of $|U_{\widetilde{\gamma}\nu}|$ and
$M_1$. In the first constraint in eq.~(\ref{eq:constraint1}) though, the 
dependence on $|U_{\widetilde{\gamma}\nu}|$ drops out.

\subsection{Constraints from the Neutrino Mass Matrix}

Further constraints appear from neutrino physics, controlled by the
BRpV parameters $\Lambda_i$ and $\epsilon_i$, and by MSSM parameters like
gaugino and higgsino masses. We do a scan over parameter space looking for 
good solutions to neutrino observables. The range in which we vary the 
parameters is given in Table \ref{tab1}.
\begin{center}
\begin{table}
\begin{tabular*}{0.5\textwidth}{@{\extracolsep{\fill}} c c c }
\hline\hline
SUSY parameter & Scanned range & Units \\ \hline
$\tan\beta$ & $[2,50]$ & $-$ \\
$|\mu|$ & $[0,1000]$ & GeV\\
$M_2$ & $2M_1$ & GeV\\
$M_1$ & $100,300,500$ & GeV\\
$m_h$ & $[114,140]$ & GeV\\
$m_A$ & $[50,6000]$ & GeV\\
$Q$ & $951.7$ & $-$\\
\hline
RpV parameter & &\\
\hline
$\epsilon_1$ & $[-1,1]$ & GeV\\
$\epsilon_2$ & $[-1,1]$ & GeV\\
$\epsilon_3$ & $[-1,1]$ & GeV\\
$\Lambda_1$ & $[-1,1]$ & GeV$^2$\\
$\Lambda_2$ & $[-1,1]$ & GeV$^2$\\
$\Lambda_3$ & $[-1,1]$ & GeV$^2$\\
\hline\hline
\end{tabular*}
\caption{Scanned ranges for PSS and RpV parameters.
\label{tab1}}
\end{table}
\end{center}
We define a $\chi^2$ value for each point in parameter space as follows
\begin{equation}
\chi^2=\left(\frac{10^3\Delta
m_{\rm{atm}}^2-2.4}{0.4}\right)^2+\left(\frac{10^5\Delta
m_{\rm{sol}}^2-7.7}{0.6}\right)^2+\left(\frac{\sin^2
\theta_{\rm{atm}}-0.505}{0.165}\right)^2+\left(\frac{\sin^2
\theta_{\rm{sol}}-0.33}{0.07}\right)^2
\end{equation}
allowing a $3\sigma$ deviation \cite{Maltoni:2004ei}. The point is accepted
if $\chi^2<4$, plus the additional condition that the reactor angle satisfies the bound $\sin^2\theta_{\rm reac}<0.05$. Since
$\tau_{3/2}$ depends directly on $|U_{\widetilde{\gamma}\nu}|$ we determined 
its maximal and minimal values 
for a given $M_1$ compatible with neutrino physics.
The numerical results are given in Table \ref{tab2}.
\begin{center}
\begin{table}[hbt]
\begin{tabular*}{0.4\textwidth}{@{\extracolsep{\fill}} c c c }
\hline\hline
$M_1$ & $|U_{\widetilde{\gamma}\nu}|^2$(min) & $|U_{\widetilde{\gamma}\nu}|^2$(max)
\\ \hline
100 GeV & $2\times10^{-16}$ & $4\times10^{-13}$ \\
300 GeV & $2\times10^{-17}$ & $3\times10^{-14}$ \\
500 GeV & $1\times10^{-17}$ & $1\times10^{-14}$ \\
\hline\hline
\end{tabular*}
\caption{Maximal and minimal values of $|U_{\widetilde{\gamma}\nu}|^2$,
consistent with neutrino experiments, for three different values of $M_1$.
\label{tab2}}
\end{table}
\end{center}
One sees that the range of possible values for $|U_{\widetilde{\gamma}\nu}|^2$
depends on the value of $M_1$, with the maximal value being around 3 orders of
magnitude greater than the minimum value for each case. Since the gravitino
lifetime $\tau_{3/2}^{-1}=
\Gamma(\widetilde{G}\rightarrow\gamma\nu)+\Gamma(\mathrm{3-body})$
depends on $|U_{\widetilde{\gamma}\nu}|^2$, $m_{3/2}$, and $M_1$, this imposes
two extra constraints in the plane $m_{3/2}-\tau_{3/2}$ that complement the one
in eq.~(\ref{eq:constraint1}).

\section{Combined constraints and results}

\begin{figure}[hbt]
  \centering
\subfloat[\it Allowed region for $M_1=100$ GeV.]{\label{r100p}
\includegraphics[angle=270,width=0.48\textwidth]{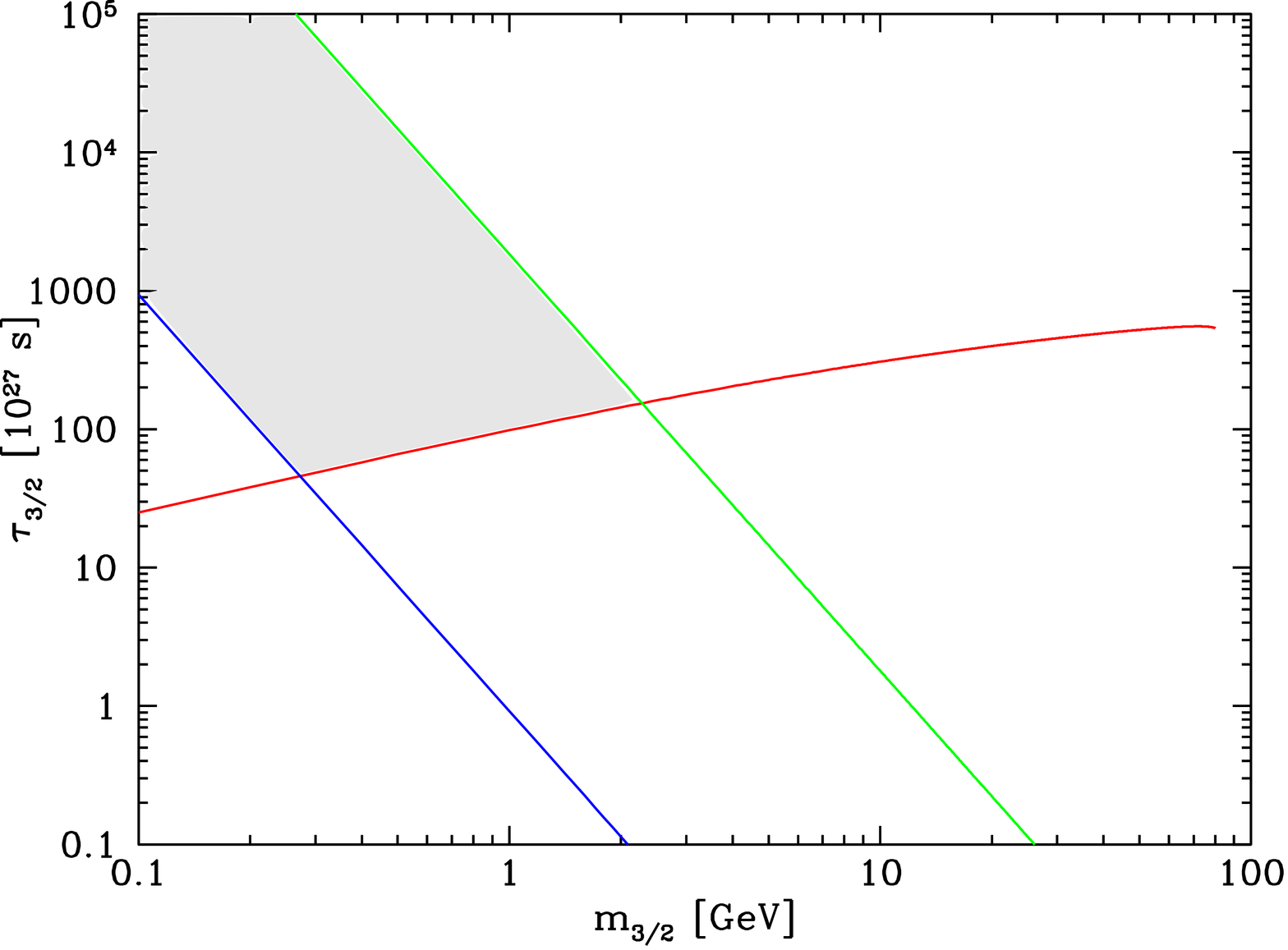}}
\hspace{0.2cm}
\subfloat[\it
Allowed region for $M_1=500$ GeV. ]{\label{r500p}
\includegraphics[angle=270,width=0.48\textwidth]{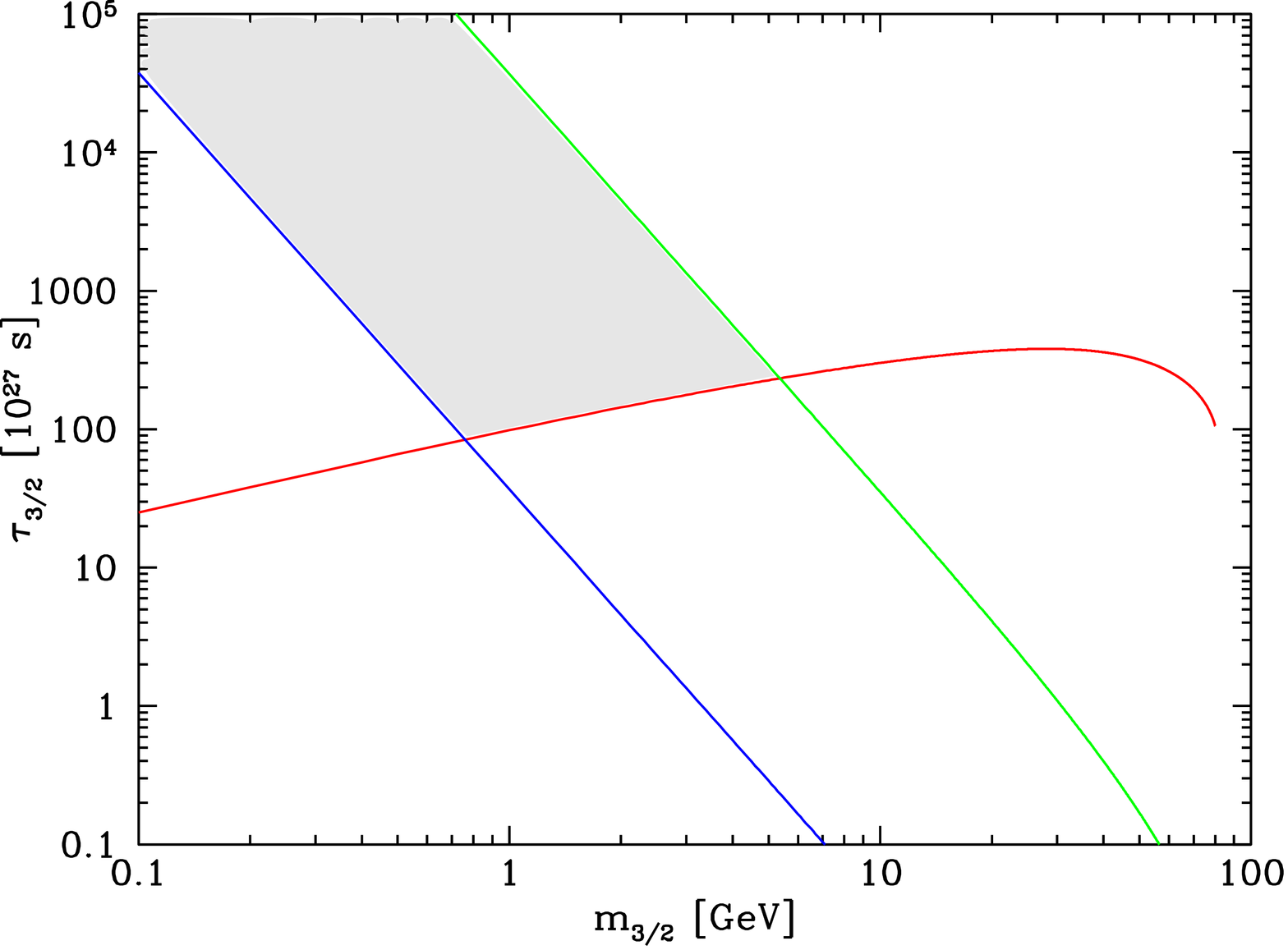}}
\caption{\it Allowed (shaded) region in the $m_{3/2}-\tau_{3/2}$ plane. The
region
above the nearly horizontal (red) line is 
allowed by the constraint in eq.~(\ref{eq:constraint1}). The region between 
the oblique lines is allowed by minimum (green) and maximum (blue) values of
$|U_{\widetilde{\gamma}\nu}|^2$.}
\label{gl}
\end{figure}

The combination of the constraints found in the previous section defines an
allowed region in the $m_{3/2}-\tau_{3/2}$ plane. This region is shown in
figs.~(\ref{r100p}) and (\ref{r500p}) for the gaugino mass,
$M_1=100$, and $500$ GeV. One sees that the constraint from the photon
spectrum, when taken together with the maximal value of
$|U_{\widetilde{\gamma}\nu}|^2$ consistent with neutrino experiments, gives a
lower bound on the gravitino lifetime. In all cases that we studied, this bound is several
orders of magnitude larger than the age of the universe, compatible with a good
candidate for dark matter. Even more interestingly, we see from these graphs
that the constraint from the photon spectrum, when combined with the minimum
allowed value of $|U_{\widetilde{\gamma}\nu}|^2$, imposes an upper bound on the
gravitino mass. This bound is near 2, 4, and 5 GeV for $M_1=100$, $300$, and
$500$ GeV.

We also note from these graphs how the 3-body decays of the gravitino become more important as $M_1$ increases. In particular, the constraint coming from the photon spectrum analysis becomes less stringent as the gravitino mass gets closer to $m_W$, which is quite evident for the case of $M_1=500$ GeV. This is expected from the fact that the strength of the gravitino photon line is proportional to the branching ratio of the 2-body decay. Conversely, for a gravitino of mass below $\sim$10 GeV this branching ratio is close to 1, and so from eq.~(\ref{eq:2body}) we see that the lifetime is approximately proportional to $m_{3/2}^{-3}$, as can be noted in the graphs.
%
%

We stress that this analysis assumes that $m_{3/2}<m_W$. For a gravitino with a
mass greater than the $Z$ boson mass the two-body decays
$\widetilde{G}\rightarrow Z\nu$ and $\widetilde{G}\rightarrow W\ell$ will be
kinematically allowed. As shown in ref.~\cite{Ibarra:2007wg}, the branching
ratio of the decay $\widetilde{G}\rightarrow\gamma\nu$ becomes very small for a
gravitino mass above $100$ GeV, and so the monochromatic line in the photon
spectrum becomes less important. A more detailed analysis of the photon spectrum
produced by the decay of the gravitino would be required in that case, which
would need to include the fragmentation of the $W$ and $Z$ bosons in addition to
the contribution from the 3-body decays.

\section{Summary}
It is explained how PSS in combination with RpV
allows to generate the neutrino masses and mixings
at the one loop level.
Then it is investigated whether in this model the
gravitino still is a good dark matter candidate.
In order to do this the gravitino decay rates into
two and three-body states are calculated.
In the three-body decay corrections are found to previous 
calculations. Since for relatively small gravitino masses
the decay is dominated by the two-body decay, the
corrections in the three-body case do not affect the final result.
Those rates allow to calculate the additional
photon flux that is induced by the gravitino decay.
Comparison of this photon flux with recent data
from the Fermi LAT collaboration, allows to put restrictions
on our model in the gravitino decay process.

Finally, combining the restrictions obtained in the neutrino
sector with the restrictions in the gravitino dark matter sector
of the same model,
an upper limit on the gravitino mass $m_{3/2}<6$ GeV is found.
The exact values of the maximal $m_{3/2}$ 
and the minimal $\tau_{3/2}$ in our model are given
in table \ref{tab:3}.
One observes a relatively weak $M_1$ dependence.
\begin{center}
\begin{table} [hbt]
\begin{tabular*}{0.4\textwidth}{@{\extracolsep{\fill}} c c c }
\hline\hline
$M_1$ [GeV] & $\tau_{3/2}$(min) [s] & $m_{3/2}$(max) [GeV]
\\ \hline
$100$ & $4.7\times10^{28}$ & $2.3$ \\
$300$ & $7.2\times10^{28}$ & $4.4$ \\
$500$ & $8.6\times10^{28}$ & $5.3$ \\
\hline\hline
\end{tabular*}
\caption{Minimal values of $\tau_{3/2}$ and maximal values of
$m_{3/2}$ as a function of $M_1$.\label{tab:3}}
\end{table}
\end{center}
It is interesting to note that most direct dark matter search
experiments disfavor the typical heavy dark matter particle 
that appears in R-Parity conserving
supersymmetric models \cite{Ahmed:2009zw,Aprile:2011hi}.
The bound, however, is not even closely as tight for a
light dark matter gravitino as it is found here.
On the other hand, with the derived limit, the model
turns out to be directly testable. For instance, if 
a relatively heavy dark matter particle is found with
$m_{3/2}\gg10$ GeV, one could immediately conclude that
this model is ruled out.
This feature of testability and the possibility of falsification
can be seen as a strong advantage over
many other models.

\begin{acknowledgments}
{\small 
B.\ K.\ has been funded by PSD73/2006. M.\ A.\ D.\ was supported by 
Fondecyt  Regular Grant \# 1100837.
Many thanks to Drs.\ Ki-Young Choi, D.\ Restrepo, and C.\ Yaguna for providing
a detailed comparison of the 3-body results.}
\end{acknowledgments}

\section{Appendix: Three-body decays formulas}

We label the five relevant diagrams with indices 1 to 5 as indicated in the text. 
For the photon and $Z$ mediated diagrams we define the invariant masses 
$s:=(k_1+k_2)^2$, and $t:=(k_1+q)^2$, where $k_1$, $k_2$ and $q$ are the 
4-momenta of the fermion, antifermion and neutrino, respectively. For the 
spin-averaged squared amplitudes and interferences we find (we define 
$m\equiv m_{3/2}$ for simplicity),
\begin{equation} \label{eq:m11}
\langle|\mathcal{M}_{1}|^2\rangle=\frac{1}{4}\left(\frac{q_f^2|U_{\widetilde{\gamma}\nu_i}|^2}{16M_P^2}\right)\frac{1}{s^2}~T_{11},
\end{equation}
\begin{equation}
\langle|\mathcal{M}_2|^2\rangle=\frac{1}{4}\left(\frac{g^2|U_{\widetilde{Z}\nu_i}|^2}{16c_W^2M_P^2}\right)\frac{1}{(s-m_Z^2)^2+m_Z^2\Gamma_Z^2}~T_{22},
\end{equation}
\begin{equation}
\langle|\mathcal{M}_3|^2\rangle=\frac{1}{4}\left(\frac{g^4\Lambda_i^2}{64\mu^2c_W^4M_P^2}\right)\frac{1}{(s-m_Z^2)^2+m_Z^2\Gamma_Z^2}~{T}_{33},
\end{equation}
\begin{equation}
2\Re\langle\mathcal{M}_{1}^{*}\mathcal{M}_2\rangle=\frac{2}{4}\left(\frac{gq_fU_{\widetilde{\gamma}\nu_i}U_{\widetilde{Z}\nu_i}}{16c_WM_P^2}\right)\frac{(s-m_Z^2)}{s\left[(s-m_Z^2)^2+m_Z^2\Gamma_Z^2\right]}~{T}_{12},
\end{equation}
\begin{equation}
2\Re\langle\mathcal{M}_{1}^{*}\mathcal{M}_3\rangle=\frac{2}{4}\left(\frac{g^2q_fU_{\widetilde{\gamma}\nu_i}\Lambda_i}{32\mu c_W^2M_P^2}\right)\frac{(s-m_Z^2)}{s\left[(s-m_Z^2)^2+m_Z^2\Gamma_Z^2\right]}~{T}_{13},
\end{equation}
\begin{equation}
2\Re\langle\mathcal{M}_2^{*}\mathcal{M}_3\rangle=\frac{2}{4}\left(\frac{g^3U_{\widetilde{Z}\nu_i}\Lambda_i}{32\mu c_W^3M_P^2}\right)\frac{1}{(s-m_Z^2)^2+m_Z^2\Gamma_Z^2}~T_{23}.
\end{equation}

For the $W$-mediated diagrams, we define $s$ and $t$ as above, with $k_1$, $k_2$ and $q$ the 4-momenta of the neutrino, antilepton and lepton, respectively. We find,
\begin{equation}
\langle|\mathcal{M}_4|^2\rangle=\frac{1}{4}\left(\frac{g^2|U_{\widetilde{W}\ell'_i}|^2}{128M_P^2}\right)\frac{1}{(s-m_W^2)^2+m_W^2\Gamma_W^2}~T_{44},
\end{equation}
\begin{equation}
\langle|\mathcal{M}_5|^2\rangle=\frac{1}{4}\left(\frac{g^4\Lambda_i^2}{256\mu^2M_P^2}\right)\frac{1}{(s-m_W^2)^2+m_W^2\Gamma_W^2}~{T}_{55},
\end{equation}
\begin{equation}
2\Re\langle\mathcal{M}_4^{*}\mathcal{M}_5\rangle=\frac{2}{4}\left(\frac{g^3U_{\widetilde{W}\ell'_i}\Lambda_i}{128\sqrt{2}\mu M_P^2}\right)\frac{1}{(s-m_W^2)^2+m_W^2\Gamma_W^2}~T_{45}.
\end{equation}

Finally, for the interference terms between these two groups of diagrams (when $f=\ell'=\ell$), we define $s$ and $t$ as above, with $k_1$, $k_2$ and $q$ the 4-momenta of the lepton, antilepton and neutrino, respectively. We find,
\begin{equation}
2\Re\langle\mathcal{M}_1^{*}\mathcal{M}_4\rangle=\frac{2}{4}\left(\frac{gq_fU_{\widetilde{\gamma}\nu_i}U_{\widetilde{W}\ell_i}}{32\sqrt{2}M_P^2}\right)\frac{(m^2-(s+t)-m_W^2)}{s\left[(m^2-(s+t)-m_W^2)^2+m_W^2\Gamma_W^2\right]}~T_{14},
\end{equation}
\begin{equation}
2\Re\langle\mathcal{M}_2^{*}\mathcal{M}_4\rangle=\frac{2}{4}\left(\frac{g^2U_{\widetilde{Z}\nu_i}U_{\widetilde{W}\ell_i}}{32\sqrt{2}c_WM_P^2}\right)\frac{(s-m_Z^2)(m^2-(s+t)-m_W^2)+m_Zm_W\Gamma_Z\Gamma_W}{\left[(s-m_Z^2)^2+m_Z^2\Gamma_Z^2\right]\left[(m^2-(s+t)-m_W^2)^2+m_W^2\Gamma_W^2\right]}~T_{24},
\end{equation}
\begin{equation}
2\Re\langle\mathcal{M}_3^{*}\mathcal{M}_4\rangle=\frac{2}{4}\left(\frac{g^3U_{\widetilde{W}\ell_i}\Lambda_i}{64\sqrt{2}\mu c_W^2M_P^2}\right)\frac{(s-m_Z^2)(m^2-(s+t)-m_W^2)+m_Zm_W\Gamma_Z\Gamma_W}{\left[(s-m_Z^2)^2+m_Z^2\Gamma_Z^2\right]\left[(m^2-(s+t)-m_W^2)^2+m_W^2\Gamma_W^2\right]}~T_{34},
\end{equation}
\begin{equation}
2\Re\langle\mathcal{M}_1^{*}\mathcal{M}_5\rangle=\frac{2}{4}\left(\frac{g^2q_fU_{\widetilde{\gamma}\nu_i}\Lambda_i}{64\mu M_P^2}\right)\frac{(m^2-(s+t)-m_W^2)}{s\left[(m^2-(s+t)-m_W^2)^2+m_W^2\Gamma_W^2\right]}~T_{15},
\end{equation}
\begin{equation}
2\Re\langle\mathcal{M}_2^{*}\mathcal{M}_5\rangle=\frac{2}{4}\left(\frac{g^3U_{\widetilde{Z}\nu_i}\Lambda_i}{64\mu c_WM_P^2}\right)\frac{(s-m_Z^2)(m^2-(s+t)-m_W^2)+m_Zm_W\Gamma_Z\Gamma_W}{\left[(s-m_Z^2)^2+m_Z^2\Gamma_Z^2\right]\left[(m^2+-(s+t)-m_W^2)^2+m_W^2\Gamma_W^2\right]}~T_{25},
\end{equation}
\begin{equation} \label{eq:m35}
2\Re\langle\mathcal{M}_3^{*}\mathcal{M}_5\rangle=\frac{2}{4}\left(\frac{g^4\Lambda_i^2}{128\mu^2c_W^2M_P^2}\right)\frac{(s-m_Z^2)(m^2-(s+t)-m_W^2)+m_Zm_W\Gamma_Z\Gamma_W}{\left[(s-m_Z^2)^2+m_Z^2\Gamma_Z^2\right]\left[(m^2-(s+t)-m_W^2)^2+m_W^2\Gamma_W^2\right]}~T_{35}.
\end{equation}

\bigskip

The total amplitude of the 3-body decays is given by the sum of all these terms, each being summed over all the relevant flavors and colors of the final states. The traces in equations (\ref{eq:m11}) to (\ref{eq:m35}) are given by,
\begin{equation}
\begin{split}
T_{11}=&~\frac{64}{3m^2}s\Big\{3m^6-3m^4(s+2t)+m^2(s^2+8st+6t^2)-s(s^2+2t(s+t))\Big\},
\end{split}
\end{equation}
\begin{equation}
\begin{split}
T_{22}=&~\frac{64}{3m^2}(c_V^2+c_A^2)s\Big\{3m^6-3m^4(s+2t)+m^2(s^2+6t^2+8st)-s(s^2+2t(s+t))\Big\},
\end{split}
\end{equation}
\begin{equation} \label{eq:t33}
\begin{split}
T_{33}=&~\frac{64}{3m^2}\Big\{(c_V^2+c_A^2)(m^2-s)(m^2(2s+t)-t(s+t))-2c_Vc_Am^2s(m^2-s-2t)\Big\},
\end{split}
\end{equation}
\begin{equation}
\begin{split}
T_{12}=&~\frac{64}{3m^2}c_Vs\Big\{ 3m^6-3m^4(s+2t)+m^2(s^2+8st+6t^2)-s(s^2+2t(s+t))\Big\},
\end{split}
\end{equation}
\begin{equation}
\begin{split}
T_{13}=&~\frac{32}{3m}s\Big\{c_V(3m^4-2m^2(s+t)-s^2+2st+2t^2)-c_A(3m^2-s)(m^2-s-2t)\Big\},
\end{split}
\end{equation}
\begin{equation}
\begin{split}
T_{23}=&~\frac{32}{3m}s\Big\{(c_V^2+c_A^2)(3m^4-2m^2(s+t)-s^2+2t^2+2st)\\
&-2c_Vc_A(3m^4-2m^2(2s+3t)+s^2+2st)\Big\},
\end{split}
\end{equation}
\begin{equation}
\begin{split}
T_{44}=&~\frac{128}{3m^2}s\Big\{3m^6-3m^4(s+2t)+m^2(s^2+8st+6t^2)-s(s^2+2st+2t^2)\Big\},
\end{split}
\end{equation}
\begin{equation}
\begin{split}
T_{55}=&~\frac{128}{3m^2}\Big\{m^4(3s+t)-m^2(s+t)(3s+t)+st(s+t)\Big\},
\end{split}
\end{equation}
\begin{equation}
\begin{split}
T_{45}=&~\frac{128}{3m}s\Big\{3m^4-m^2(3s+4t)+t(2s+t)\Big\},
\end{split}
\end{equation}
\begin{equation}
\begin{split}
T_{14}=&~\frac{64}{3m^2}s\Big\{3m^6-m^4(3s+7t)+m^2t(4s+5t)-t^2(s+t)\Big\},
\end{split}
\end{equation}
\begin{equation}
\begin{split}
T_{24}=&~\frac{64}{3m^2}(c_V-c_A)s\Big\{3m^6-m^4(3s+7t)+m^2t(4s+5t)-t^2(s+t)\Big\},
\end{split}
\end{equation}
\begin{equation}
\begin{split}
T_{34}=&~\frac{64}{3m^2}(c_V-c_A)\Big\{m^5(3s+t)-m^3(3s^2+6st+2t^2)+mt(s+t)(2s+t)\Big\},
\end{split}
\end{equation}
\begin{equation}
\begin{split}
T_{15}=&~\frac{64}{3m}s\Big\{3m^4-m^2(3s+4t)+t(2s+t)\Big\},
\end{split}
\end{equation}
\begin{equation}
\begin{split}
T_{25}=&~\frac{64}{3m}(c_V-c_A)s\Big\{3m^4-m^2(3s+4t)+t(2s+t)\Big\},
\end{split}
\end{equation}
\begin{equation}
\begin{split}
T_{35}=&~\frac{64}{3m^2}(c_V-c_A)\Big\{(m^2-s)(m^2s+(m^2-t)(s+t))+m^2s(m^2-s-2t)\Big\}.
\end{split}
\end{equation}

Finally those amplitudes are applied to the
golden rule for decays in order to obtain the 
partial and total decay rate for each process
$\tilde G \rightarrow 2 + 3 + 4$
\begin{equation}
 d\Gamma=|{\mathcal{M}}|^2
\frac{S}{2 m_{3/2}}
\left[ 
\frac{d^3 p_2}{(2\pi)^3 2 E_2}
\frac{d^3 p_3}{(2\pi)^3 2 E_3}
\frac{d^3 p_4}{(2\pi)^3 2 E_4}
\right]
(2\pi)^4 \delta^4(p_1-p_2-p_3-p_4)\quad.
\end{equation}


\end{document}